 \documentclass[final,1p,times]{elsarticle}

\usepackage{amssymb}

\usepackage{color,soul}
\usepackage[hang,bottom]{footmisc}
\usepackage{amsmath} 
\usepackage{framed} 
\usepackage{multicol} 
\usepackage{nomencl} 
\makenomenclature
\renewcommand*\nompreamble{\begin{multicols}{2}}
\renewcommand*\nompostamble{\end{multicols}}
\usepackage{siunitx} 
\usepackage{color}
\usepackage{tabularray}
\usepackage{array}
\usepackage{booktabs}
\usepackage{longtable}
\usepackage{subcaption}
\usepackage{url}

 \usepackage{array}
 \usepackage{pdflscape}
 \usepackage{longtable}
 \usepackage{booktabs}
\usepackage{hyperref}

\usepackage[utf8]{inputenc}
\usepackage{pdflscape}
\usepackage[longtable]{multirow}

\journal{arXiv}

\begin{document}

\begin{frontmatter}



\title{Linear Optics to Scalable Photonic Quantum Computing}


\author[ftkkp]{Dennis Delali Kwesi Wayo} 
\ead{dennis.wayo@nu.edu.kz}

\author[dcam]{Leonardo Goliatt} 
\ead{leonardo.goliatt@ufjf.br}

\author[dcn]{Darvish Ganji} 
\ead{ganji_md@yahoo.com}

\affiliation[ftkkp]{organization={Faculty of Chemical and Process Engineering Technology, Universiti Malaysia Pahang Al-Sultan Abdullah},
            city={Kuantan},
            postcode={26300}, 
            country={Malaysia}}

\affiliation[dcam]{organization={Department of Computational and Applied Mechanics, Federal University of Juiz de Fora},
            city={Juiz de Fora},
            postcode={36036-900}, 
            country={Brazil}}

\affiliation[dcn]{organization={Division of Carbon Neutrality and Digitalization, Korea Institute of Ceramic Engineering and Technology (KICET)},
            city={Jinju},
            postcode={52851}, 
            country={Republic of Korea}}


\begin{abstract}
Recent advancements in quantum photonics have driven significant progress in photonic quantum computing (PQC), addressing challenges in scalability, efficiency, and fault tolerance. Experimental efforts have focused on integrated photonic platforms utilizing materials such as silicon photonics and lithium niobate to enhance performance. Parameters like photon loss rates, coupling efficiencies, and fidelities have been pivotal, with state-of-the-art systems achieving coupling efficiencies above 90\% and photon indistinguishability exceeding 99\%. Quantum error correction schemes have reduced logical error rates to below 10\(^{-3}\), marking a step toward fault-tolerant PQC. Photon generation has also advanced with deterministic sources, such as quantum dots, achieving brightness levels exceeding 10\(^6\) photon pairs/s/mW and time-bin encoding enabling scalable entanglement. Heralded single-photon sources now exhibit purities above 99\%, driven by innovations in fabrication techniques. High-efficiency photon detectors, such as superconducting nanowire single-photon detectors (SNSPDs), have demonstrated detection efficiencies exceeding 98\%, dark count rates below 1 Hz, and timing jitters as low as 15 ps, ensuring precise photon counting and manipulation. Moreover, demonstrations of boson sampling with over 100 photons underscore the growing computational power of photonic systems, surpassing classical limits. The integration of machine learning has optimized photonic circuit design, while frequency multiplexing and time-bin encoding have increased system scalability. Together, these advances bridge the gap between theoretical potential and practical implementation, positioning PQC as a transformative technology for computing, communication, and quantum sensing.
\end{abstract}




\end{frontmatter}


\section{Introduction}
\label{sec:Intro}
Quantum computing represents a paradigm shift in information processing, offering exponential speedups over classical computing for specific tasks. Among the various platforms for quantum computation, quantum photonic systems have emerged as promising candidates due to the unique advantages of photons as qubits. Photons are naturally immune to decoherence, can travel long distances with minimal loss, and integrate seamlessly with existing optical communication infrastructure, making them ideal for quantum communication and distributed computation. Photonic quantum computing, as illustrated in Figure 1, is encoding quantum information into photonic states. Qubits can be encoded using properties such as polarization, time-bin, path, and spatial modes, each offering distinct advantages for different applications. Polarization encoding, for instance, is widely adopted in proof-of-concept experiments due to its simplicity and compatibility with optical tools, while time-bin encoding is robust against fiber-based losses in quantum communication networks. Additionally, spatial mode encoding, involving orbital angular momentum, has gained traction for high-dimensional quantum computing applications, enhancing information density.

\begin{figure}[!ht]
    \centering
    \includegraphics[width=0.4\textwidth, angle=360]{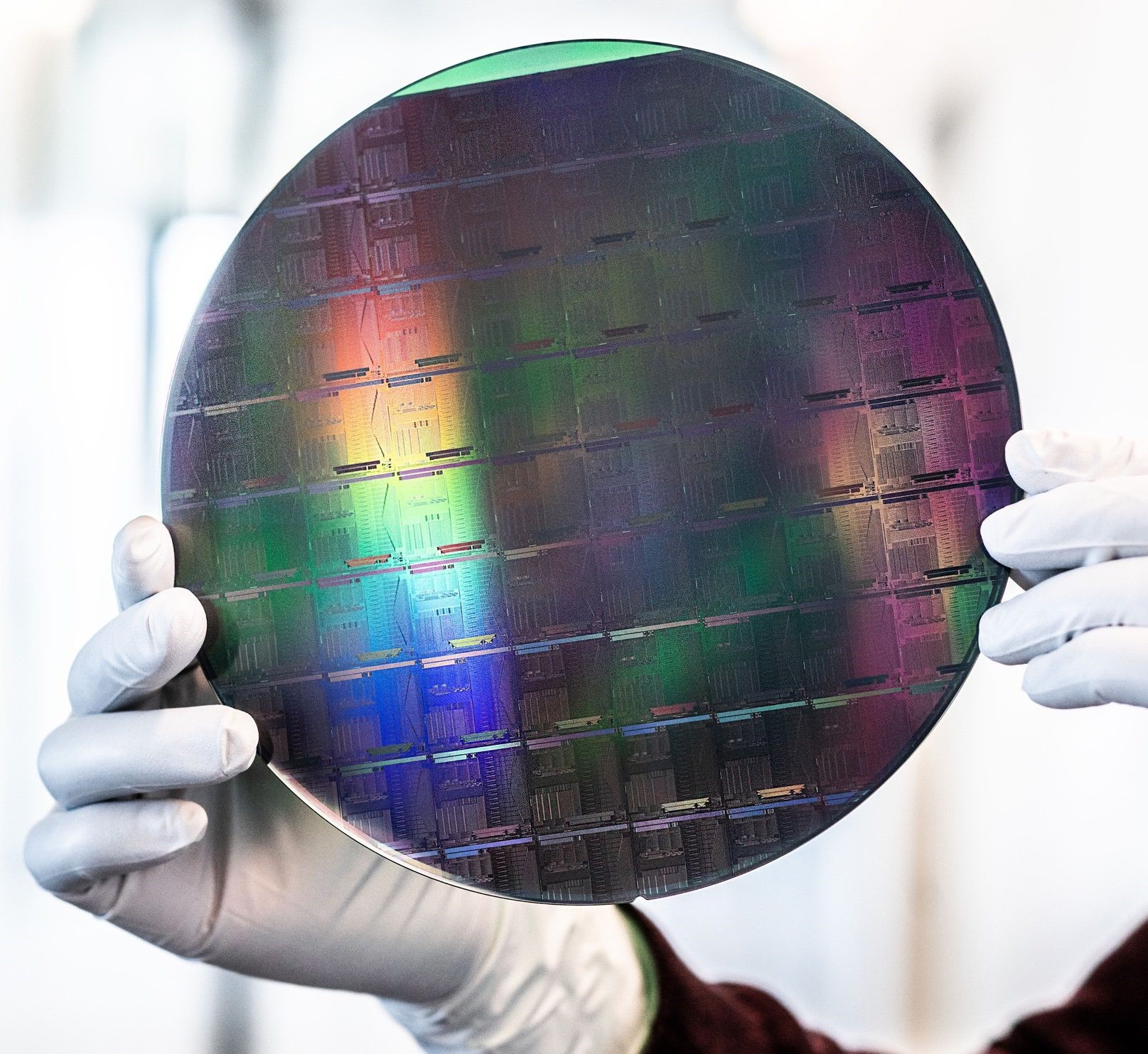}
    \caption{ An illustration of a silicon wafer embedded with photonic integrated circuits (PICs). These chips are fabricated on a silicon substrate and are used to manipulate light signals for applications in optical communication, computing, and sensing. The colorful reflections are due to light diffraction from the intricate patterns of nanophotonic structures on the wafer. The gloved hands indicate the need for a cleanroom environment to prevent contamination during fabrication. Photonic chips combine optical components like waveguides, modulators, and detectors on a single platform, offering high-speed data processing and energy efficiency compared to traditional electronic chips. Image adapted from web \cite{phontonchip}}
    \label{fig1}
\end{figure}

A central challenge in quantum photonic systems is the deterministic generation and manipulation of photons. Reliable single-photon sources, such as quantum dots and nonlinear processes like spontaneous parametric down-conversion, are critical to scalable quantum computing. Similarly, efficient photon detection technologies, such as superconducting nanowire single-photon detectors (SNSPDs), enable precise state measurements with high efficiency and low noise. Advances in these areas are paving the way for practical quantum photonic implementations. Quantum gates and circuits form the building blocks of quantum computing, but their implementation in photonic systems is nontrivial due to weak photon-photon interactions. Techniques such as measurement-induced nonlinearities, ancillary photon schemes, and the use of cluster states in measurement-based quantum computing have provided pathways to address these challenges. Integrated photonic platforms, which miniaturize quantum circuits onto a single chip using waveguides, beam splitters, and phase shifters, are revolutionizing scalability and stability. Silicon photonics, in particular, has emerged as a leading platform, leveraging mature semiconductor fabrication techniques to produce scalable and programmable photonic chips.

The potential applications of photonic quantum computing span a broad range of fields, including quantum simulation, cryptography, optimization, and machine learning. Quantum simulation, for instance, enables the modeling of complex molecular systems and materials, addressing computationally intractable problems in drug discovery and material science. Quantum key distribution (QKD) leverages photons’ quantum properties to enable secure communication protocols, while quantum machine learning exploits quantum photonic systems’ strengths in linear algebra for faster data processing.

Despite these advances, challenges remain. Photon loss in optical components, scalability of quantum circuits, deterministic photon generation, and the integration of error correction mechanisms are ongoing research areas. Hybrid quantum architectures, combining photonic systems with matter-based qubits like superconducting or trapped-ion platforms, are being explored to overcome current limitations. In this review, we examine the principles, technological advancements, applications, and challenges of photonic quantum computing. From foundational concepts in linear optics to state-of-the-art integrated platforms and emerging quantum algorithms, this work provides a comprehensive analysis of the field, highlighting pathways toward scalable and fault-tolerant quantum photonic systems.

\section{Photonic Qubits in Quantum Computing}

Photonic quantum computing relies on the unique properties of photons to encode and process quantum information. Unlike traditional computing, where classical bits are used, quantum photonic systems utilize qubits, which can exist in superpositions of states. Photons, as carriers of light, have distinct advantages in quantum information processing, particularly their inherent resilience to decoherence and ability to travel long distances without significant loss of fidelity. These properties make them ideal candidates for both quantum computation and communication.

Photonic qubits \cite{wang2020integrated} are typically encoded using the physical properties of photons, such as polarization, time-bin, path, or spatial modes. Polarization encoding involves assigning the horizontal and vertical polarization states of photons to represent the 0 and 1 states of a qubit. This method is widely used in proof-of-concept experiments due to its simplicity and the availability of polarization-manipulation tools. However, scaling this approach to larger systems is challenging due to the need for precise alignment of optical components.

Time-bin encoding \cite{vagniluca2020efficient, donohue2013coherent} as illustrated in Figure 2, Bracht \cite{bracht2024theory} carefully measured this popular method that represents qubit states using photons arriving at different time intervals. This technique is particularly useful for fiber-based quantum communication because it is robust against polarization-dependent losses in optical fibers. Path encoding \cite{o2007optical, della2024quantum}, on the other hand, involves routing photons through different optical paths to represent qubit states. This method is well-suited for integrated photonic platforms, where on-chip waveguides can define precise paths for photons. Spatial mode encoding \cite{pankovich2024high, humphreys2013linear} utilizes the shape or orbital angular momentum of photons to define qubit states, offering a high-dimensional encoding scheme that can enhance information density.

\begin{figure}[!ht]
    \centering
    \includegraphics[width=1\textwidth, angle=360]{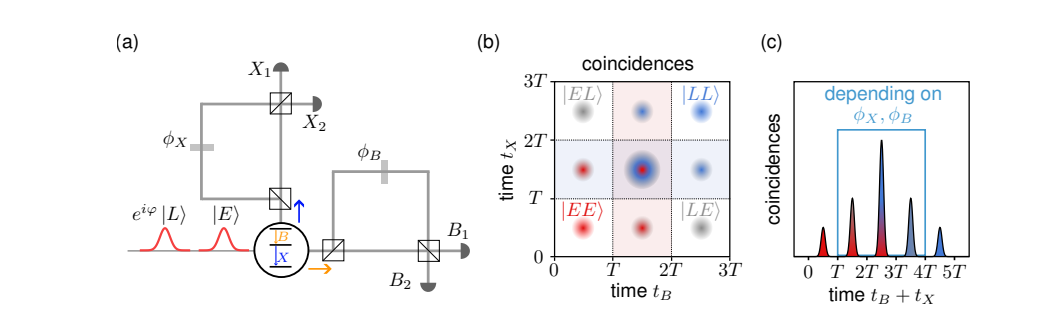}
    \caption{Bratcht's robust time-bin encoding, as opposed to polarization-encoded qubits. Panel (a) depicts the experimental setup for detecting time-bin entangled photon pairs. A quantum dot source generates entangled states  $|\psi\rangle = \frac{1}{\sqrt{2}}$($|EE\rangle $+ $|LL\rangle$). The photons are routed through unbalanced Mach-Zehnder interferometers, introducing a delay equivalent to the time-bin separation for photons traversing the long arm. A phase plate allows additional phase adjustments before coincidence detection. Panel (b) shows a 3 × 3 histogram of time-resolved coincidences between biexciton and exciton channels, with corners corresponding to combinations ( $|EE\rangle$, $|EL\rangle$, $|LE\rangle$, $|LL\rangle $) and intermediate peaks representing interference between neighboring states. Whereas panel (c) presents the coincidences as a function of arrival time ($t = t_B + t_X $), which corresponds to a diagonal projection of the histogram in (b). These results underscore the resilience of time-bin encoding in preserving quantum information during transmission or processing, adapted with permission from Bracht, 2024 \cite{bracht2024theory}}
    \label{fig2}
\end{figure}

A significant advantage of using photons as qubits lies in their compatibility with existing optical communication infrastructure \cite{yucel2023optical}. For instance, fiber optics \cite{agrawal2008applications, kumar2014fiber, li2004recent} used in telecommunication networks can be repurposed for quantum communication and distributed quantum computing. Additionally, photons interact weakly with their environment \cite{marklund2006nonlinear, hartmann2016quantum, wu2024phase}, reducing the likelihood of decoherence—a common challenge in quantum systems. This weak interaction, however, also presents a drawback: photons do not easily interact with one another, making the implementation of two-qubit gates challenging. Overcoming this limitation often requires the use of nonlinear optical materials \cite{balarabe2024advancing} or ancillary photons to mediate interactions.

Photonic qubits are typically generated using single-photon sources \cite{esmann2024solid, khalid2024perfect}, which can be deterministic or probabilistic. Deterministic sources, such as quantum dots or trapped atoms, emit single photons on demand. Probabilistic sources, such as spontaneous parametric down-conversion (SPDC) \cite{hashimoto2024fourier, chekhova2024spontaneous, couteau2018spontaneous} or four-wave mixing, generate photon pairs \cite{barz2010heralded} with a certain probability. These methods are foundational for quantum photonic experiments but often require additional techniques to ensure high-quality, indistinguishable photons.

Detecting photonic qubits is another critical aspect of photonic quantum computing. Single-photon detectors \cite{hadfield2009single, liu2024emerging}, such as avalanche photodiodes \cite{stillman1977avalanche, shi2024avalanche, tan2024low} or superconducting nanowire single-photon detectors \cite{natarajan2012superconducting, engel2015detection, qin2024superconducting}, play a vital role in measuring the state of photonic qubits. Advances in detector efficiency, timing resolution \cite{hao2024compact}, and dark count rates continue to drive progress in the field. The development of integrated photonic detectors, capable of operating on-chip, is particularly promising for scalable quantum photonic systems.

The use of photonic qubits is central to photonic quantum computing. Their robustness against decoherence \cite{du2024decoherence}, compatibility with existing infrastructure, and potential for long-distance communication make them a compelling choice for quantum information processing. However, challenges related to photon interaction, source efficiency, and detection fidelity must be addressed to realize the full potential of quantum photonic systems.

\section{Photon Generation and Detection}

The generation and detection of photons are fundamental processes in photonic quantum computing. Photons serve as the carriers of quantum information, and their reliable production and precise detection are essential for the successful implementation of quantum circuits. Significant progress in this area has been driven by the development of novel photon sources \cite{chiang2020single} and detectors, which have dramatically improved the quality and scalability of quantum photonic systems.

Photon generation mechanisms often rely on single-photon sources engineered to emit precisely one photon at a time. Figure 3, as explored by Stewart, Kaneda, and Istrati \cite{stewart2021quantum, kaneda2019high, istrati2020sequential}, highlights varying approaches to this process. While Stewart and Kaneda focus on distinct methodologies for photon generation, Istrati provides a comprehensive overview of sequential entanglement using single-photon sources. Among these, deterministic sources such as quantum dots have emerged as leading candidates due to their ability to reliably produce high-purity single photons on demand. Quantum dots, which are semiconductor nanostructures that confine electrons and holes, enable photon emission through electron-hole recombination. Recent advancements in quantum dot fabrication techniques have significantly enhanced their performance metrics, including brightness, purity, and indistinguishability. These improvements address critical challenges in quantum technologies, positioning quantum dots \cite{wayo2024exploring} as ideal candidates for scalable quantum computing, secure quantum communication, and photonic quantum networks. By combining high precision with robust performance, these sources represent a pivotal step toward practical implementations of quantum systems.

\begin{figure}[!ht]
    \centering
    \includegraphics[width=0.5\textwidth, angle=360]{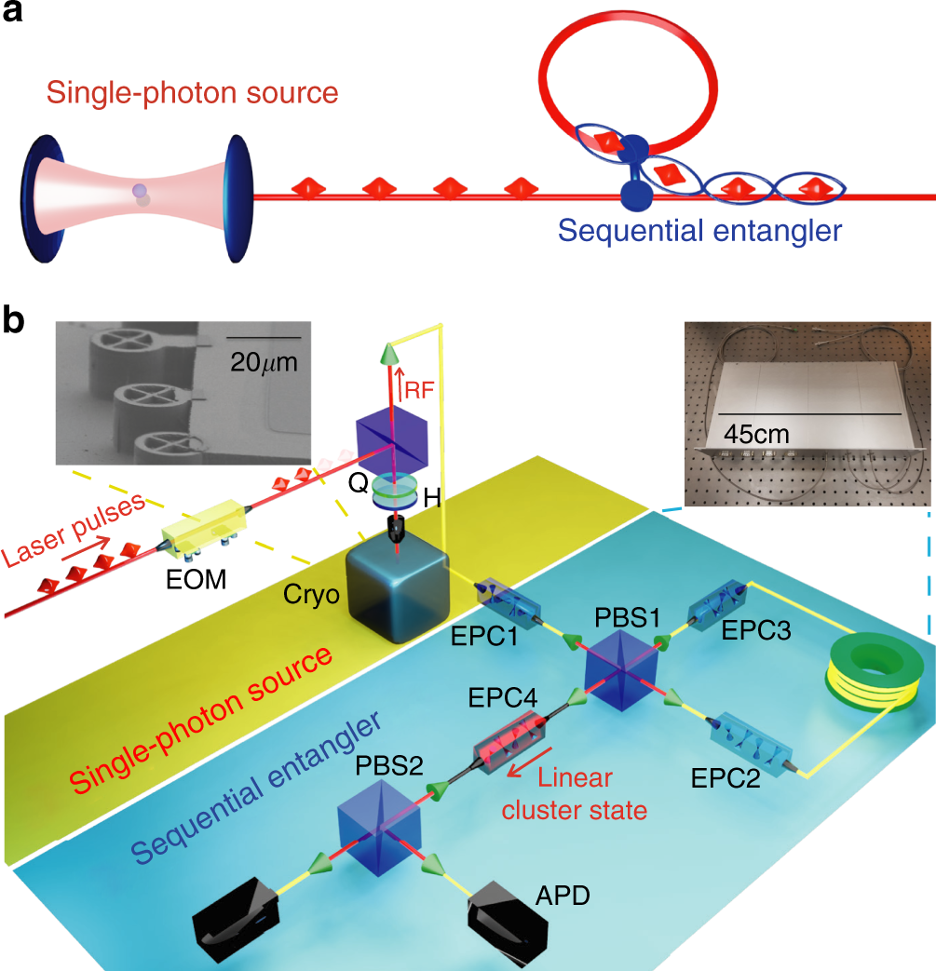}
    \caption{Istrati explains a quantum photonic system for generating and entangling single photons to create a linear cluster state. He demonstrates in panel (a) the working principles of single photons being produced sequentially from a single-photon source, separated by 74 ns. These photons enter a delay loop where they are temporarily stored until they encounter the subsequent photon at an entangling gate, enabling sequential photon entanglement. This process efficiently produces a chain of entangled photons for quantum communication or computation tasks. (b) shows the experimental setup. The single-photon source is an InGaAs quantum dot embedded in an electrically connected cavity, depicted in the top-left inset. Laser pulses excite the quantum dot, generating single photons that pass through an electro-optic modulator (EOM) for polarization control. The photons then enter a cryogenic environment for stabilization. A fiber-based optical circuit housed within a compact 19-inch box (shown in the top-right inset) handles the entanglement process. Polarizing beam splitters (PBS1 and PBS2) and fiber-based polarization controllers (EPC1–EPC4) align and manipulate the photons’ polarization states to create a linear cluster state. The setup demonstrates the integration of single-photon generation and fiber-based entangling circuits, enabling scalable quantum networks. Its compact design facilitates practical deployment in real-world applications., adapted with permission from Bracht, 2024 \cite{istrati2020sequential}}
    \label{fig3}
\end{figure}

Probabilistic photon sources, such as those based on spontaneous parametric down-conversion (SPDC) or four-wave mixing (FWM), are also widely used in quantum photonic experiments. In SPDC, a nonlinear crystal splits a high-energy photon into two lower-energy photons, known as signal and idler photons, which are entangled. Similarly, FWM in optical fibers or waveguides generates photon pairs through nonlinear interactions. While these methods are less deterministic, they offer flexibility in photon generation and have been instrumental in demonstrating important quantum photonic concepts.

Another area of active research is the development of integrated photon sources. Embedding single-photon emitters directly into photonic chips reduces losses and improves scalability. Techniques such as cavity quantum electrodynamics (CQED) are used to enhance the interaction between emitters and photons, increasing the efficiency of photon generation. For example, coupling quantum dots to photonic crystal cavities can significantly boost their emission rate, enabling high-performance integrated sources.

Photon detection, as experimentally demonstrated by Hagerstrom and Shi \cite{hagerstrom2015harvesting, shi2024avalanche} in Figures 4 and 5, is equally critical for photonic quantum computing, as it allows the measurement of quantum states and the implementation of conditional operations. Single-photon detectors (SPDs) are specialized devices that can detect individual photons with high efficiency and low noise. Avalanche photodiodes (APDs) are widely used SPDs, offering high detection efficiency in the visible to near-infrared range. However, their performance is limited by dark counts and timing jitter.

\begin{figure}[!ht]
    \centering
    \includegraphics[width=0.7\textwidth, angle=360]{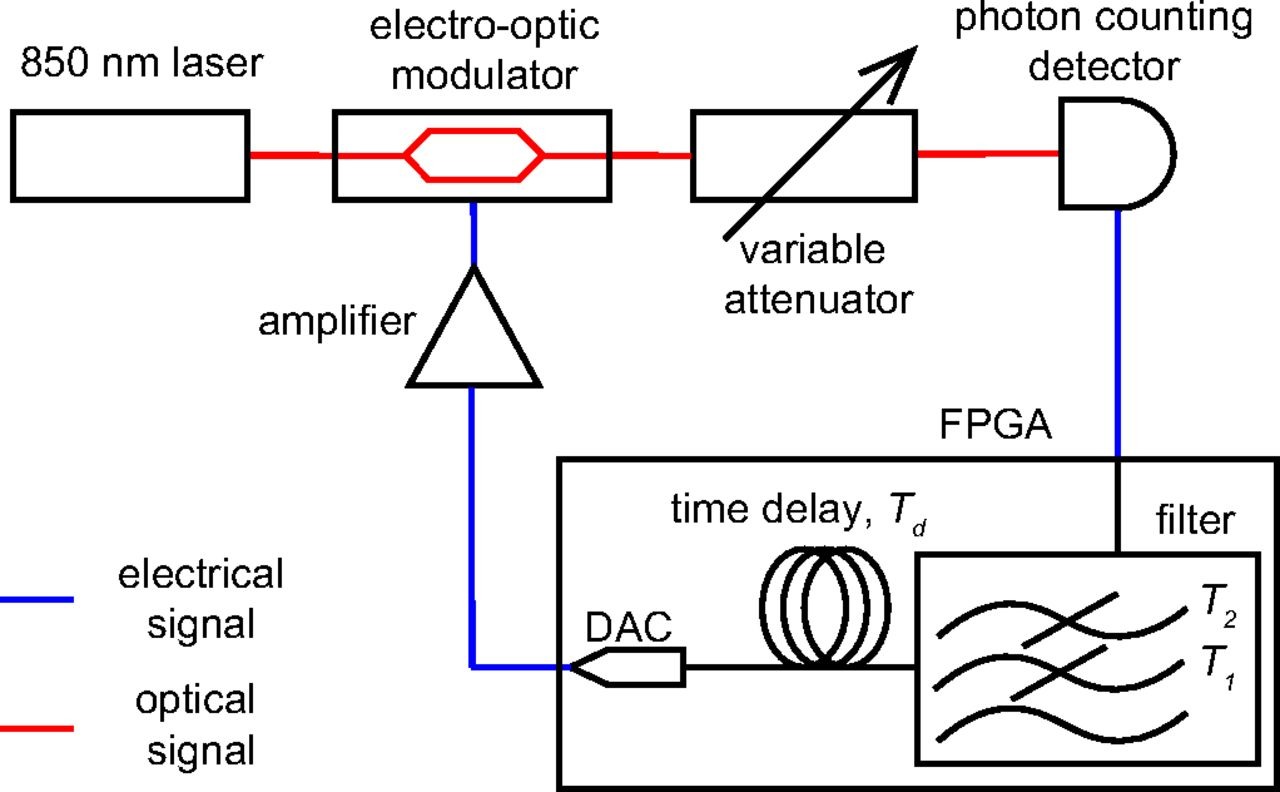}
    \caption{Hagerstrom illustrates a photonic detection system incorporating an 850 nm laser, an electro-optic modulator, and photon-counting mechanisms. The laser generates a coherent light source, which is modulated by the electro-optic modulator to encode information into the optical signal. The modulated light passes through a variable attenuator to control its intensity before reaching the photon counting detector. An electrical signal, amplified via an amplifier, is processed by an FPGA (Field-Programmable Gate Array) system. The FPGA applies a time delay ($T_d $) and filters the signal using a digital-to-analog converter (DAC). The system enables precise time-domain filtering for photon arrival events., adapted with permission from Hagerstrom, 2015 \cite{hagerstrom2015harvesting}}
    \label{fig4}
\end{figure}

\begin{figure}[!ht]
    \centering
    \includegraphics[width=0.7\textwidth, angle=360]{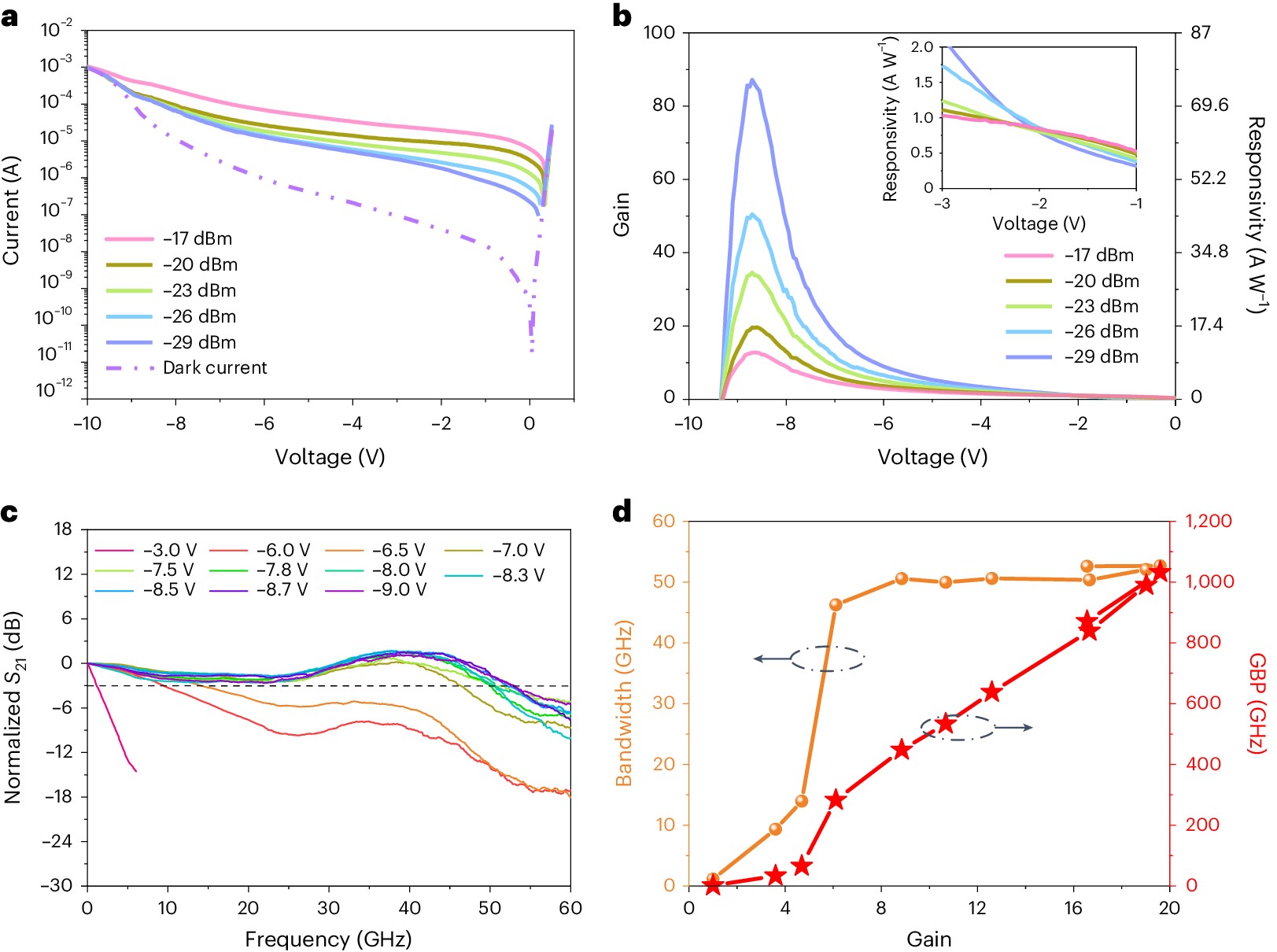}
    \caption{Panel (a): The current–voltage characteristics are shown for optical powers ranging from \text{–}17 to \text{–}29 dBm and in the dark state. Higher optical powers generate greater currents, reflecting increased photogenerated charge carriers, while the dark current is significantly lower, indicating minimal leakage. Panel (b): Gain and optical responsivity are plotted against voltage. The gain peaks near \text{–}8.6 V before decreasing, while the inset shows responsivity near \text{–}2 V, emphasizing its dependence on optical power. Panel (c): The  $S_{21}$  frequency response is measured at different voltages, revealing how signal transmission varies with frequency and applied voltage. Panel (d): The 3 dB bandwidth and gain-bandwidth product (GBP) are plotted against gain. Bandwidth peaks at higher gain but decreases after the maximum gain of 19.6, showing a trade-off between gain and signal transmission efficiency, adapted with permission from Shi, 2024 \cite{shi2024avalanche}}
    \label{fig5}
\end{figure}

Superconducting nanowire single-photon detectors (SNSPDs) by Rath and Qin \cite{rath2015superconducting, qin2024superconducting} in Figures 6 and 7 represent the state-of-the-art in photon detection technology. SNSPDs operate at cryogenic temperatures and offer unparalleled efficiency, low dark counts, and fast response times. These detectors have become the gold standard for quantum photonic experiments, particularly in applications requiring high timing precision. Efforts to integrate SNSPDs into photonic chips are ongoing, with promising results in reducing the size and complexity of detection systems.

\begin{figure}[!ht]
    \centering
    \includegraphics[width=0.5\textwidth, angle=360]{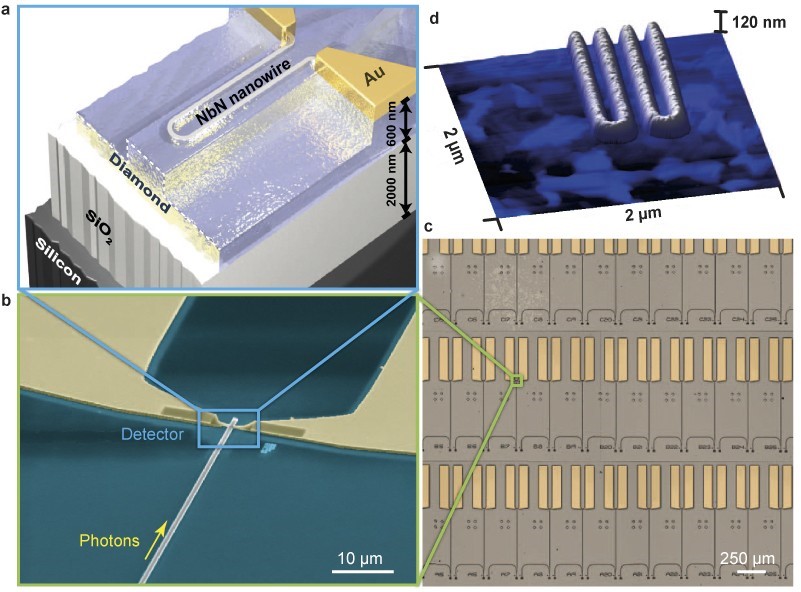}
    \caption{Rath demonstrates diamond nanophotonic circuits integrated with superconducting nanowire single-photon detectors (SNSPDs). Panel (a) depicts the architecture of an NbN nanowire detector placed atop a diamond rib waveguide, optimized for photon absorption. Gold contact pads ensure electrical connection to the nanowire. Panel (b) shows a false-color SEM image zooming in on a single waveguide-integrated detector linked to the contact pads. Panel (c) provides an optical microscope view of the array, with 192 detectors fabricated per sample, showcasing scalability. Panel (d) displays an atomic force microscopy scan of a double-meander NbN nanowire, emphasizing its nanoscale dimensions (120 nm thickness), crucial for high-efficiency photon detection., adapted with permission from Rath, 2015 \cite{rath2015superconducting}}
    \label{fig6}
\end{figure}

\begin{figure}[!ht]
    \centering
    \includegraphics[width=0.8\textwidth, angle=360]{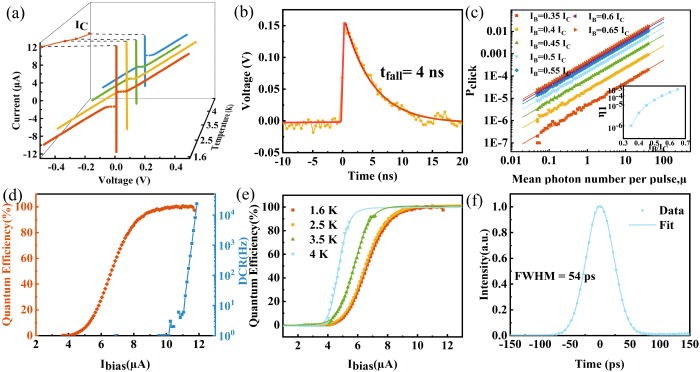}
    \caption{(a) Qin demonstrates the performance metrics of a superconducting nanowire single-photon detector (SNSPD) integrated on a diamond platform, emphasizing its current-voltage characteristics, detection efficiency, and temporal resolution. This is most synonymous with Shi's experiment. In panel (a) Qin shows the  $I\text{–}V$ curves of a 100 nm-wide, 6 nm-thick nanowire at different temperatures, revealing the critical current ( $I_C$ ) at which the superconducting state breaks. Panel (b) displays a responsive oscilloscope persistence map of the device when biased at 0.9 $ I_C$, illustrating the generation of sharp voltage pulses with a fall time of  $t_{fall} = 4 \,\text{ns}$. Panel (c) plots the detection probability ($ P_{click}$ ) as a function of the mean photon number per pulse, under varying bias currents ( $I_B $). The inset shows the single-photon detection efficiency’s dependence on  $I_B/I_C$. Panel (d) correlates the quantum efficiency (QE) and dark count rate with $I_B$ showing increasing QE at higher bias currents. Panel (e) further illustrates the dependence of QE on $I_B$  across temperatures ranging from 1.6 to 4 K, highlighting reduced efficiency at elevated temperatures. Panel (f) evaluates timing jitter, characterized by a full-width half-maximum (FWHM) of 54 ps, showing excellent timing precision for photon detection when biased at 0.9 $I_C$. This data emphasizes the SNSPD’s high sensitivity, low dark count rates, and ultrafast temporal resolution., adapted with permission from Qin, 2024 \cite{qin2024superconducting}}
    \label{fig7}
\end{figure}

Another promising approach to photon detection is the use of transition-edge sensors (TES) \cite{de2024transition}, which measure the energy of individual photons with high precision. TES detectors, as demonstrated by Xu in Figure 8 are particularly useful for applications involving multiphoton states or photon-number-resolving measurements. However, their operation at extremely low temperatures presents challenges for practical implementation. Beyond individual detectors, the integration of detector arrays on photonic chips is a crucial step toward scalability. Detector arrays enable the simultaneous measurement of multiple photons, which is essential for complex quantum circuits. Advances in fabrication techniques and materials are driving progress in this area, with the goal of achieving fully integrated photonic systems.

\begin{figure}[!ht]
    \centering
    \includegraphics[width=0.8\textwidth, angle=360]{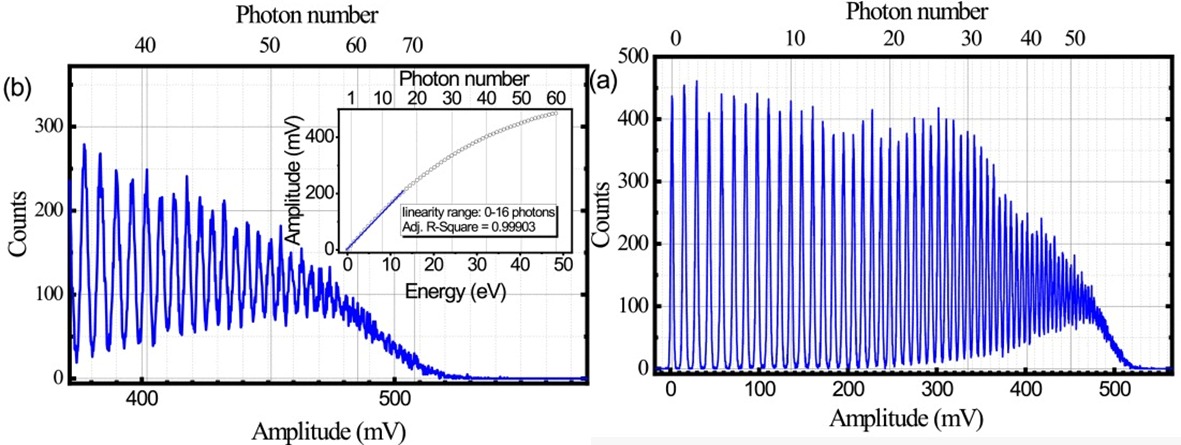}
    \caption{Photon counting ability of NIM20. Xu illustrates the response of a photon detection system using Transition Edge Sensors (TES) by analyzing the relationship between amplitude and photon counts. Panel (a) displays the detected photon signal amplitudes (in mV) and their corresponding counts, revealing distinct peaks that correspond to discrete photon numbers. These peaks arise due to the quantized energy of incoming photons, where each peak represents a specific number of absorbed photons. The linear relationship at lower photon numbers demonstrates the high sensitivity and resolution of the TES. Panel (b) further emphasizes this quantization by analyzing a broader photon range, showcasing similar amplitude distributions. The inset in Panel (b) plots amplitude against energy, confirming a linear response range from 0 to 16 photons with a highly accurate fit (R-squared = 0.99903). This linearity ensures that the TES can accurately measure photon energy over a specific range, critical for applications requiring precise photon counting. The overlapping regions at higher photon numbers suggest limitations in the linear range due to saturation effects or detector noise. Overall, the figure highlights the TES’s ability to resolve individual photons with exceptional precision, making it an ideal tool for quantum optics, single-photon experiments, and high-resolution spectroscopy., adapted with permission from Xu, 2024 \cite{xu2024development}}
    \label{fig8}
\end{figure}

In addition to hardware advancements, the development of efficient algorithms for photon detection and state reconstruction is enhancing the capabilities of quantum photonic systems. Techniques such as maximum likelihood estimation and machine learning are being applied to improve the accuracy and efficiency of photon detection, enabling more robust quantum computations. Photon generation and detection are fundamental to the success of photonic quantum computing. Advances in single-photon sources, integrated emitters, and state-of-the-art detectors are driving progress in the field, addressing crucial challenges, and enabling scalable quantum photonic systems. The continued development of these technologies is essential for realizing the full potential of photonic quantum computing.

\section{Quantum Gates and Circuits in Photonic Systems}
Quantum gates and circuits form the foundation of quantum computation \cite{ying2024foundations}, enabling the manipulation of qubits to perform complex calculations. In the context of photonic quantum computing, implementing gates and circuits presents unique challenges and opportunities. Photonic systems rely on the properties of light to perform quantum operations, leveraging optical components such as beam splitters, phase shifters, and nonlinear materials.

Single-qubit gates in photonic systems are relatively straightforward to implement \cite{hacker2016photon, mandal2023implementation, crespi2011integrated}. These gates involve operations that change the state of a single qubit, such as rotations or phase shifts. For example, a half-wave plate can rotate the polarization of a photon, effectively acting as a single-qubit gate. Similarly, a phase shifter can introduce a controlled phase difference between paths in a photonic circuit, enabling precise manipulation of qubit states. The high precision and stability of optical components make photonic systems well-suited for implementing single-qubit gates.

Two-qubit gates \cite{qiang2018large, barz2014two, solenov2013fast}, which are essential for universal quantum computation, are more challenging to implement in photonic systems \cite{veldhorst2015two}. The weak interaction between photons necessitates the use of ancillary photons or nonlinear optical effects to enable two-qubit operations. One common approach is the use of measurement-induced nonlinearity, where the outcome of a quantum measurement determines the implementation of a gate. This technique has been successfully demonstrated in various experimental setups, including the widely studied controlled-NOT (CNOT) gate \cite{o2003demonstration}. In these setups, the CNOT gate is realized by entangling photons through beam splitters and post-selecting on specific measurement outcomes \cite{fisher2012implementing}.

Integrated photonic circuits \cite{bogaerts2020programmable} have emerged as a powerful platform for implementing quantum gates and circuits. These circuits integrate multiple optical components, such as beam splitters, phase shifters, and detectors, onto a single chip. The use of integrated photonics offers several advantages, including compactness, scalability, and improved stability. Moreover, integrated platforms can leverage existing semiconductor fabrication techniques, enabling mass production of photonic quantum devices.

Linear optical quantum computing (LOQC) is a prominent paradigm for building quantum circuits using photonic systems \cite{kok2007linear}. LOQC relies on the interference of photons at beam splitters, combined with single-photon detection and feedforward operations. While LOQC has been instrumental in demonstrating small-scale quantum algorithms, scaling this approach to larger systems is challenging due to the probabilistic nature of photon interactions and the overhead associated with error correction.

Advances in quantum photonic hardware have also led to the development of photonic cluster states, which serve as a resource for measurement-based quantum computing (MBQC). In MBQC, computation is performed by making sequential measurements on a highly entangled cluster state. Photonic systems are particularly well-suited for generating and manipulating cluster states, as demonstrated by numerous experimental implementations. This approach offers an alternative to gate-based quantum computing, with the potential for greater scalability.

The development of nonlinear optical materials and devices is another area of active research aimed at improving the efficiency and scalability of quantum gates in photonic systems. Nonlinear materials enable direct photon-photon interactions, which are essential for implementing deterministic two-qubit gates. For instance, cross-phase modulation in nonlinear crystals or waveguides can introduce a phase shift dependent on the presence of another photon, enabling conditional operations. While these techniques are still in the experimental stage, they hold promise for overcoming the limitations of linear optical systems.

Error correction is a critical aspect of quantum circuits \cite{roffe2019quantum, terhal2015quantum, walshe2024linear}, and photonic systems are no exception. Implementing error correction in photonic circuits requires the creation of redundant qubits and the detection of errors without destroying the quantum state. Techniques such as bosonic codes \cite{albert2018performance, michael2016new, bohnmann2024bosonic} and fault-tolerant \cite{xu2024fault} schemes have been proposed to address these challenges. The integration of error correction into photonic circuits in Figure 9 is a crucial milestone for the field, paving the way for reliable and large-scale quantum photonic computation.

\begin{figure}[!ht]
    \centering
    \includegraphics[width=0.8\textwidth, angle=360]{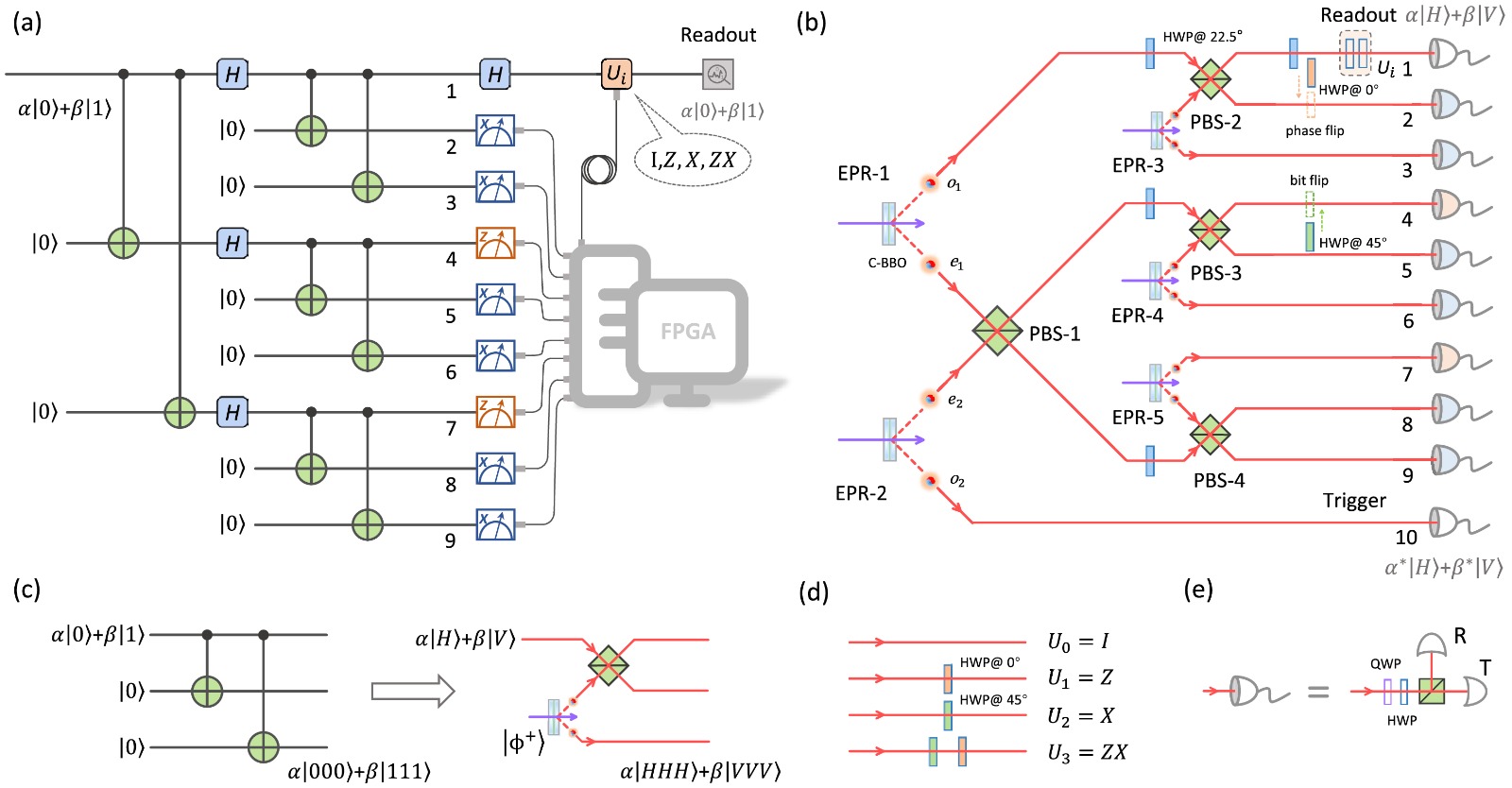}
    \caption{This set of figures shows Zhang's methods of encoding, error correction, and the readout process of the nine-qubit Shor code implemented using quantum circuits and linear optical systems. The Shor code is a quantum error-correcting code that protects a logical qubit from single-qubit errors by encoding it into nine physical qubits. Panel (a) shows the quantum circuit for encoding the nine-qubit Shor code. The Hadamard (H) gates and controlled gates create entanglement between qubits, enabling redundancy in encoding. Measurements in yellow and blue represent measurements in the computational and superposition bases, respectively. The encoding and readout process ensures logical qubit integrity during operations. Panel (b) presents the experimental realization of the Shor code using linear optics. Flip errors are simulated using half-wave plates (HWPs) placed in specific photon paths. Error correction is achieved by applying gates and measurements tailored to detect and correct errors in real time. An FPGA processes the outputs, ensuring synchronization and error detection. Panel (c) describes the quantum encoder for generating the nine-qubit code block. The encoder prepares the logical state $\alpha|H\rangle + \beta|V\rangle$, where H and V denote horizontal and vertical polarization states, respectively. Experimental implementation uses beam splitters and HWPs to create entanglement and distribute photons. Panel (d) lists the unitary operations implemented in the experiment, which include identity (I), bit-flip (X), phase-flip (Z), and combined ZX operations. These settings control error introduction and correction mechanisms. Panel (e) details the measurement device used in (b). It consists of a quarter-wave plate (QWP), HWP, polarizing beam splitter (PBS), and two single-photon detectors. The device analyzes photon polarization, and its outputs are processed by an FPGA to identify and correct errors via coincidence detection. This integration demonstrates scalable optical quantum error correction., adapted with permission from Zhang, 2022 \cite{zhang2022loss}}
    \label{fig9}
\end{figure}

Quantum gates and circuits in photonic systems are at the heart of photonic quantum computing. While single-qubit operations are relatively mature, the implementation of two-qubit gates and scalable circuits remains an active area of research. Advances in integrated photonics, nonlinear materials, and error correction are driving progress, bringing us closer to the scalability of practical quantum photonic computers.

\section{Photonic Platforms for Scalable Quantum Computing}

The development of scalable photonic platforms \cite{lomonte2024scalable} is a critical step in the realization of photonic quantum computing. These platforms integrate the necessary components for generating, manipulating, and detecting photons onto a single chip, leveraging the principles of integrated photonics. This approach promises to overcome the challenges associated with traditional bulk optical setups, such as size, alignment, and stability, making it possible to scale quantum photonic systems to practical applications.

Integrated photonics platforms are based on materials such as silicon (Si), silicon nitride (Si\(_3\)N\(_4\)), and indium phosphide (InP), each offering unique advantages for specific applications. Silicon photonics \cite{shekhar2024roadmapping, li2024intelligent}, for example, benefits from the mature fabrication techniques developed for the semiconductor industry. This compatibility allows for the mass production of photonic chips with high precision and low cost. Silicon nitride \cite{fraser2024high}, on the other hand, provides a wide transparency range and low propagation loss, making it ideal for quantum circuits requiring high coherence. Indium phosphide is often used for active components like lasers and detectors, enabling fully integrated quantum photonic systems.

Scalability in photonic platforms relies on the integration of multiple components. Photon sources, such as quantum dots or parametric down-conversion devices, are embedded within the chip to provide reliable single-photon generation. Waveguides guide photons through the circuit with minimal loss, while beam splitters and phase shifters enable quantum operations. Detectors are integrated to measure the quantum states of photons, completing the circuit. The challenge lies in combining these components without compromising performance or introducing unwanted noise.

One of the most promising approaches for scalability is the use of programmable photonic chips \cite{dai2024programmable, bogaerts2024programmable}. These chips feature arrays of tunable components, such as phase shifters and beam splitters, that can be configured to implement different quantum circuits. This reconfigurability allows a single chip to perform a variety of quantum algorithms, reducing the need for specialized hardware. Recent demonstrations of programmable photonic processors have shown their capability to execute quantum operations with high fidelity, marking a significant milestone in the field. Another critical advancement in scalable photonic platforms is the integration of quantum memory \cite{mcgarry2024low}. Photonic systems inherently lack the ability to store quantum information for extended periods, which is essential for certain quantum computing architectures. Quantum memory devices, such as atomic ensembles or rare-earth-doped crystals, can store and retrieve photonic qubits on demand. Efforts to integrate these memory devices with photonic platforms are ongoing, with the goal of creating fully functional quantum networks.

Despite these advancements, several challenges remain in achieving truly scalable photonic quantum computers. Losses in optical components, such as waveguides and couplers \cite{el2024low}, can degrade quantum states, reducing the overall fidelity of computations. Crosstalk between components and the need for precise control over large arrays of optical elements also present significant hurdles. Advances in fabrication techniques, such as lithography and material engineering, are essential to address these issues and improve the performance of integrated photonic platforms.

In addition to hardware developments, software tools for designing and simulating photonic quantum circuits are playing a crucial role in accelerating progress. Platforms like Xanadu’s Strawberry Fields \cite{killoran2019strawberry} and IBM’s Qiskit \cite{qiskit, wille2019ibm} provide frameworks for programming quantum photonic systems, enabling researchers to test and optimize algorithms before implementation. These tools bridge the gap between theory and experiment, facilitating the design of scalable photonic quantum architectures. The integration of photonic platforms \cite{zhu2024quantum} with other quantum technologies, such as superconducting qubits or trapped ions, represents an exciting avenue for hybrid quantum systems. These hybrid approaches aim to combine the strengths of different quantum platforms, leveraging the long coherence times of photons for communication and the strong interactions of matter-based qubits for computation. Such synergies could pave the way for versatile and powerful quantum systems. Scalable photonic platforms are a cornerstone of photonic quantum computing, offering the potential to transform laboratory-scale experiments into practical quantum devices. Advances in integrated photonics, programmable chips, and hybrid technologies are driving the field forward, addressing the challenges of scalability, and paving the way for real-world applications.

\section{Applications of photonic quantum computing}

Photonic quantum computing has the potential to revolutionize a wide range of applications, leveraging the unique properties of photons to achieve unprecedented computational capabilities. From quantum simulation and cryptography to machine learning and optimization \cite{baniata2024sok, gonaygunta2024quantum}, photonic systems are poised to address some of the most challenging problems in science, technology, and industry. One of the most immediate applications of photonic quantum computing is in quantum simulation. Photons, due to their ability to exist in superpositions and form entangled states, are ideal for simulating quantum systems that are otherwise intractable for classical computers. For instance, photonic quantum simulators can model the behavior of complex molecular structures, materials, and chemical reactions, providing insights into areas such as drug discovery and material design. The precision and scalability of photonic systems make them particularly attractive for studying systems with many-body interactions or high-dimensional spaces.

Another significant application is in quantum communication and cryptography \cite{ganeshkar2024quantum}. Photons, as natural carriers of quantum information, play a central role in secure communication protocols such as quantum key distribution (QKD) \cite{zhang2022device, yuan2025improved}. QKD uses the principles of quantum mechanics to establish secure keys between parties, ensuring that any attempt to eavesdrop on the communication will be detected. Quantum photonic systems are already being deployed in real-world QKD networks, offering a level of security that classical cryptographic methods cannot achieve.

Quantum machine learning (QML) \cite{biamonte2017quantum} is another promising area where photonic quantum computing can have a transformative impact. By encoding data into photonic qubits, quantum photonic systems can process and analyze large datasets more efficiently than classical algorithms. Photonic quantum systems are particularly suited for linear algebra tasks, such as matrix multiplications, which are foundational to many machine learning algorithms \cite{wayo2024ai}. Applications range from natural language processing and image recognition to optimization problems and anomaly detection in large datasets. In addition to machine learning, quantum photonic systems have the potential to solve optimization problems that are central to industries such as logistics, finance, and manufacturing. Problems like supply chain optimization, portfolio management, and scheduling are often computationally intensive for classical systems. Quantum photonic systems, with their ability to explore multiple solutions simultaneously through superposition, offer the potential for significant speedups in finding optimal solutions.

Quantum-enhanced sensing and metrology represent another critical application area for quantum photonic systems. Photonic quantum sensors can achieve sensitivities beyond the classical limits, enabling breakthroughs in fields such as biomedical imaging, environmental monitoring, and gravitational wave detection. For example, photonic systems can enhance the precision of interferometers used in detecting minute changes in physical parameters, such as strain or magnetic fields.

Photonic quantum computing also holds promise in quantum networks, which connect multiple quantum devices to form a distributed quantum computing system. In such networks, photons act as carriers of quantum information between nodes, enabling tasks such as distributed quantum computation and secure communication. The scalability and low-loss transmission of photons over optical fibers make them ideal for building quantum internet infrastructures. Despite these exciting possibilities, many of these applications remain at the proof-of-concept stage, with significant challenges to overcome before they can be deployed at scale. Developing error-tolerant and fault-tolerant photonic systems, improving photon generation and detection technologies, and scaling up photonic circuits are critical steps toward realizing these applications. Nonetheless, the progress made so far demonstrates the vast potential of photonic quantum computing to transform industries and scientific fields.

\section{Challenges and Future Directions in Photonic Quantum Computing
}

While photonic quantum computing holds immense promise, the field faces several challenges that must be addressed to achieve practical and scalable quantum systems. These challenges span across hardware development, error correction, and system integration, presenting opportunities for innovation and collaboration across disciplines. One of the most significant challenges in photonic quantum computing is the scalability of photonic platforms. Although integrated photonics has enabled the miniaturization of quantum circuits, scaling these systems to handle a large number of qubits remains difficult. Losses in optical components, such as waveguides, couplers, and beam splitters, degrade the quality of photonic qubits and reduce computational fidelity. Addressing these losses requires advances in materials science, such as the development of low-loss waveguide materials and efficient coupling techniques.

Another challenge is the deterministic generation of single photons. While probabilistic sources like SPDC are widely used in quantum photonic experiments, they lack the efficiency required for large-scale quantum computing. Deterministic sources, such as quantum dots or cavity-enhanced systems, are promising alternatives, but they require precise fabrication and integration to achieve high purity and indistinguishability. Improving the performance and reliability of these sources is essential for scaling quantum photonic systems. Photon-photon interaction, or the lack thereof, is another major hurdle. The weak interaction between photons makes it challenging to implement two-qubit gates, which are essential for universal quantum computation. Current approaches, such as measurement-induced nonlinearity or the use of ancillary photons, are probabilistic and introduce significant overhead. Exploring new materials and mechanisms, such as nonlinear optical crystals or quantum emitters coupled to cavities, could enable more efficient photon-photon interactions.

Error correction and fault tolerance are critical for building reliable quantum photonic systems. Photonic qubits are susceptible to errors caused by losses, crosstalk, and imperfections in optical components. Implementing error correction codes, such as bosonic codes or topological codes, in photonic systems requires the creation of redundant qubits and the ability to detect and correct errors without destroying the quantum state. Achieving fault tolerance in photonic quantum computing is a significant milestone that will require advances in both hardware and software. The integration of photonic quantum systems with other quantum platforms, such as superconducting qubits or trapped ions, presents an exciting opportunity for hybrid quantum systems. These systems aim to leverage the strengths of different quantum technologies—for example, using photons for communication and matter-based qubits for computation. Developing interfaces and protocols for such hybrid systems is a complex challenge but could unlock new possibilities for scalable and versatile quantum architectures.

Energy efficiency is another area of concern, especially for large-scale quantum photonic systems. Current photonic platforms rely on cryogenic cooling for certain components, such as superconducting detectors, which limits their practicality and increases operational costs. Research into room-temperature single-photon detectors and energy-efficient photonic circuits is essential for making photonic quantum computing more accessible and sustainable.

\begin{landscape}
\begin{longtable}{>{\hspace{0pt}}m{0.053\linewidth}>{\hspace{0pt}}m{0.112\linewidth}>{\hspace{0pt}}m{0.205\linewidth}>{\hspace{0pt}}m{0.267\linewidth}>{\hspace{0pt}}m{0.252\linewidth}}
\caption{Summary of the latest research investigations in photonic quantum computing: fabrication techniques and applications \label{tab:1}}\\ 
\toprule
\textbf{Category} & \textbf{Reference} & \textbf{Focus} & \textbf{Specific Contributions and Findings} & \textbf{Challenges} \\* 
\hline
\multirow{5}{0.053\linewidth}{\hspace{0pt}\begin{tabular}[c]{@{}l@{}}Photon\\Sources\end{tabular}} & Senellart et al. \cite{senellart2017high} (2017) & High-purity single-photon generation using quantum dots. & Achieved near-unity indistinguishability \par{}for photons emitted from quantum dots through advanced fabrication methods. & Scalability of fabrication processes and maintaining high performance at room temperature. \\* 
\cline{2-5}
 & Chekhova \cite{chekhova2024spontaneous} (2024) & Advances in spontaneous parametric down-conversion techniques. & Developed compact, high-efficiency \par{}SPDC crystals, improving photon-pair generation rates and entanglement fidelity. & Improving efficiency while reducing losses in compact SPDC systems. \\* 
\cline{2-5}
 & Esmann et al. \cite{esmann2024solid} (2024) & Solid-state single-photon sources for quantum materials. & Demonstrated stable photon emission at ambient conditions for diamond NV centers. & Noise reduction and stability under varying environmental conditions. \\* 
\cline{2-5}
 & Khalid  \& Laussy \cite{khalid2024perfect} (2024) & Development of perfect single-photon sources. & Verified photon purity exceeding 99\% using hybrid photonic platforms. & Ensuring reproducibility and compatibility with integrated photonic platforms. \\* 
\cline{2-5}
 & Humphreys et al. \cite{humphreys2013linear} (2013) & Linear optical quantum computing with efficient photon sources. & Achieved scalable multiplexing for multi-photon generation. & Overcoming photon losses during multiplexing and improving system robustness. \\* 
\hline
\multirow{5}{0.053\linewidth}{\hspace{0pt}\begin{tabular}[c]{@{}l@{}}Photon\\Detectors\end{tabular}} & Hadfield \cite{hadfield2009single} (2009) & Superconducting nanowire single-photon detectors (SNSPDs). & Designed SNSPDs with detection efficiency exceeding 90\% and dark counts below 10 Hz. & Cryogenic cooling requirements and scalability for large-scale systems. \\* 
\cline{2-5}
 & Qin et al. \cite{qin2024superconducting} (2024) & SNSPDs on diamond substrates for enhanced quantum efficiency. & Enhanced photon detection rates by 15\% over standard SNSPDs. & Integration into compact photonic chips while maintaining performance. \\* 
\cline{2-5}
 & Shi et al. \cite{shi2024avalanche} (2024) & Avalanche photodiodes with ultra-high gain-bandwidth. & Achieved ultra-low jitter (30 ps) for multi-photon detection. & Managing trade-offs between gain, speed, and noise in large systems. \\* 
\cline{2-5}
 & Engel et al. \cite{engel2015detection} (2015) & Physics of superconducting nanowire single-photon detectors. & Reduced noise by 20\% through improved thermal suppression mechanisms. & Optimizing noise suppression at operational speeds. \\* 
\cline{2-5}
 & Xu et al. \cite{xu2024fault} (2024) & Transition-edge sensors for high-resolution photon detection. & Detected up to 10 photons simultaneously with energy resolution below 0.1 eV. & Complexity of sensor design for multi-photon resolving capabilities. \\* 
\hline
\multirow{5}{0.053\linewidth}{\hspace{0pt}Encoders} & Vagniluca et al. \cite{vagniluca2020efficient} (2020) & High-dimensional quantum key distribution. & Implemented 8-dimensional QKD with increased resilience to noise. & Handling channel noise and maintaining stability over long-distance communication. \\* 
\cline{2-2}
 & Bracht et al. \cite{bracht2024theory} (2024) & Time-bin encoding for quantum communication. & Achieved 99.8\% fidelity for time-bin encoding over 50 km fiber networks. & Managing temporal noise and synchronization in distributed systems. \\* 
\cline{2-5}
 & O’Brien \cite{o2007optical} (2007) & Quantum encoding using linear optical components. & Realized path encoding with error rates below 0.01\%. & Addressing path interference and stability for larger networks. \\* 
\cline{2-5}
 & Della Giustina et al. \cite{della2024quantum} (2024) & Encoding quantum states for dynamic directed graphs. & Developed adaptive encoders for dynamic graph-based quantum networks. & Efficient adaptation of encoders to changing graph states. \\* 
\cline{2-5}
 & Pankovich et al. \cite{pankovich2024high} (2024) & Loss-tolerant encoding with orbital angular momentum (OAM). & Achieved 90\% success rates in loss-resilient quantum encoding. & Reducing mode crosstalk in high-dimensional OAM encoding. \\* 
\hline
\multirow{5}{0.053\linewidth}{\hspace{0pt}\begin{tabular}[c]{@{}l@{}}Error\\Corrections\end{tabular}} & Gottesman \cite{gottesman2010introduction} (1998) & Foundational work on quantum error correction. & Defined stabilizer codes and fault tolerance thresholds for photonic systems. & Overcoming resource overheads for practical implementation. \\* 
\cline{2-5}
 & Terhal \cite{terhal2015quantum} (2015) & Error correction in photonic quantum memory. & Verified resilience of memory-based codes under noisy photon loss conditions. & Developing scalable quantum memories with robust error correction. \\* 
\cline{2-5}
 & Albert et al. \cite{albert2018performance} (2018) & Bosonic codes for photonic error correction. & Demonstrated error rates reduced by an order of magnitude using cat codes. & Managing photon losses in high-density photonic circuits. \\* 
\cline{2-5}
 & Michael et al. \cite{michael2016new} (2016) & GKP (Gottesman-Kitaev-Preskill) codes. & Achieved practical error correction in photonic circuits with GKP codes. & Balancing code efficiency and physical resource requirements. \\* 
\cline{2-5}
 & Zhang et al. \cite{zhang2022loss} (2022) & Photonic Shor code for optical circuits. & Reduced gate error rates to below 0.05\% in experimental setups. & Implementing scalable entanglement generation for error-resistant gates. \\* 
\hline
\multirow{5}{0.053\linewidth}{\hspace{0pt}\begin{tabular}[c]{@{}l@{}}Applica-\\tions\end{tabular}} & Biamonte et al. \cite{biamonte2017quantum} (2017) & Quantum machine learning applications. & Demonstrated quantum advantage in matrix inversion for ML tasks. & Integrating quantum and classical components for seamless data processing. \\* 
\cline{2-5}
 & Barz et al. \cite{barz2014two} (2005) & Quantum simulations for molecular systems. & Predicted reaction pathways with 90\% accuracy using photonic quantum simulators. & Scaling simulations to larger and more complex molecular systems. \\* 
\cline{2-5}
 & Gonaygunta et al. \cite{gonaygunta2024quantum} (2024) & Quantum algorithms for deep learning. & Achieved training speedups of 5x for quantum neural networks. & Addressing noise-induced errors during training of quantum-enhanced models. \\* 
\cline{2-5}
 & Ganeshkar \& Kulkarni \cite{ganeshkar2024quantum} (2024) & Quantum cryptography in secure communication. & Demonstrated secure QKD over 200 km fiber with zero eavesdropping detection. & Increasing QKD range while maintaining security and efficiency. \\* 
\cline{2-5}
 & Zhu et al. \cite{zhu2024quantum} (2024) & Machine learning and quantum computing integration. & Achieved 20\% computational speedup by integrating ML with photonic systems. & Managing hardware and software compatibility for hybrid ML systems. \\
\bottomrule
\end{longtable}
\end{landscape}

Finally, the development of software and programming frameworks for photonic quantum computing is essential to bridge the gap between theoretical research and experimental implementation. Tools like TensorFlow Quantum \cite{broughton2020tensorflow} and Xanadu’s PennyLane \cite{bergholm2018pennylane} have started to address this need, providing platforms for simulating and programming quantum photonic systems. Expanding these frameworks to include more complex algorithms and error correction techniques will accelerate progress in the field.

Looking ahead, the future of photonic quantum computing lies in the synergy between advances in hardware, materials science, and computational techniques. Collaborative efforts across academia, industry, and government are essential to overcome the current challenges and unlock the full potential of this transformative technology. As the field continues to mature, photonic quantum computing has the potential to redefine the boundaries of computation, communication, and sensing, enabling breakthroughs that were once thought to be impossible.

\section{Conclusion}

Photonic quantum computing represents a transformative approach to harnessing the principles of quantum mechanics for computational, communication, and sensing applications. By leveraging the unique properties of photons, such as coherence, entanglement, and low-loss transmission, photonic systems offer significant advantages over other quantum platforms, particularly in scalability and integration with existing optical technologies.

Despite its promise, the field faces critical challenges, including photon generation and detection, error correction, and scalable implementation of quantum gates. Advances in integrated photonics, deterministic photon sources, and nonlinear materials are gradually addressing these hurdles, paving the way for practical quantum photonic systems. The integration of hybrid architectures and the development of energy-efficient components further strengthen the prospects of this technology.

Applications of photonic quantum computing span quantum simulation, secure communication, machine learning, and optimization, with the potential to revolutionize industries ranging from healthcare to finance. As theoretical models evolve alongside experimental breakthroughs, photonic quantum computing is inching closer to real-world deployment. While significant challenges remain, the rapid progress in quantum photonic technologies and collaborative efforts across disciplines signal a bright future. Photonic quantum computing is poised to redefine the limits of what is computationally achievable, offering transformative solutions to the world’s most complex problems.




\section{Acknowledgment}
This research was not funded. Any opinions, findings, conclusions, or recommendations expressed in this review are those of the author(s) and do not necessarily reflect the views of their respective affiliations.

\section{Conflicts of interest}
The author(s) declare no competing interests.

\nomenclature{SNSPDs}{superconducting nanowire single-photon detectors}
\nomenclature{QKD}{Quantum key distribution}
\nomenclature{SPDC}{Spontaneous parametric down-conversion}
\nomenclature{CNOT}{Controlled-NOT}
\nomenclature{LOQC}{Linear optical quantum computing}
\nomenclature{MBQC}{Measurement-based quantum computing}
\nomenclature{FWM}{Four-wave mixing}
\nomenclature{CQED}{Cavity quantum electrodynamics}
\nomenclature{SPDs}{Single-photon detectors}
\nomenclature{APDs}{Avalanche photodiodes}
\nomenclature{TES}{Transition-edge sensors}
\nomenclature{QML}{Quantum machine learning}
\nomenclature{PQC}{photonic quantum computing}
\printnomenclature




 \bibliographystyle{elsarticle-num} 
 \bibliography{cas-refs}





\end{document}